\documentclass[usenatbib, useAMS, fleqn]{mn2e}
\usepackage{mathtools} 
\usepackage{amssymb} 
\usepackage{upgreek} 
\usepackage{epsfig, graphicx}
\usepackage[caption=false]{subfig}
\usepackage{times}

\newcommand{\ud}{\mathrm{d}}
\newcommand{\Msun}{\ensuremath{M_{\odot}}}

\newcommand{\Mbh}{\ensuremath{M_\bullet}}

\newcommand{\apjl}{ApJL}
\newcommand{\aj}{AJ}
\newcommand{\apj}{ApJ}
\newcommand{\mnras}{MNRAS}

\newcommand{\apjs}{ApJS}
\newcommand{\araa}{ARA\&A}
\newcommand{\aap}{A\&A}
\newcommand{\aapr}{A\&AR}
\newcommand{\nat}{Nature}
\newcommand{\pr}{Phys.~Rep.}
\newcommand{\pasj}{PASJ}
\newcommand{\ssr}{Space Sci.~Rev.}

\title[Contribution of SNRs to X-rays near SMBHs]{The contribution of young core-collapse supernova remnants to the X-ray emission near quiescent supermassive black holes}

\author[Rimoldi et al.]{A.~Rimoldi,$^{1}$\thanks{E-mail: rimoldi@strw.leidenuniv.nl} E.~M.~Rossi,$^{1}$ E. Costantini,$^{2}$ and S.~Portegies Zwart$^{1}$ \\
$^1$Leiden Observatory, Leiden University, Niels Bohrweg 2, NL-2333 CA Leiden, the Netherlands\\
$^2$SRON Netherlands Institute for Space Research, Sorbonnelaan 2, NL-3584 CA Utrecht, the Netherlands}
\date{}
\pagerange{\pageref{firstpage}--\pageref{lastpage}} \pubyear{2015}

\begin{document} 

\maketitle
\label{firstpage}

\begin{abstract}
Appreciable star formation, and, therefore, numerous massive stars, are frequently found near supermassive black holes (SMBHs).
As a result, core-collapse supernovae in these regions should also be expected.
In this paper, we consider the observational consequences of predicting the fate of supernova remnants (SNRs) in the sphere of influence of quiescent SMBHs.
We present these results in the context of `autarkic' nuclei, a model that describes quiescent nuclei as steady-state and self-sufficient environments where the SMBH accretes stellar winds with no appreciable inflow of material from beyond the sphere of influence.
These regions have properties such as gas density that scale with the mass of the SMBH.
Using predictions of the X-ray lifetimes of SNRs originating in the sphere of influence, we make estimates of the number of core collapse SNRs present at a given time.
With the knowledge of lifetimes of SNRs and their association with young stars, we predict a number of core-collapse SNRs that grows from $\sim 1$ around Milky Way-like ($4.3 \times 10^{6}~\Msun$) SMBHs to $\sim 100$ around the highest mass ($10^{10}~\Msun$) SMBHs.
The presence of young SNRs will amplify the X-ray emission near quiescent SMBHs, and we show that the total core-collapse SNR emission has the potential to influence soft X-ray searches for very low-luminosity SMBHs.
Our SNR lifetime estimates also allow us to predict star formation rates in these regions.
Assuming a steady-state replenishment of massive stars, we estimate a star formation rate density of $2 \times 10^{-4}$ $\Msun\,\mathrm{yr}^{-1}\,\mathrm{pc}^{-2}$ around the Milky Way SMBH, and a similar value around other SMBHs due to a weak dependence on SMBH mass.
This value is consistent with currently available observations.
\end{abstract}


\section{Introduction} \label{sec:introduction}
Supermassive black holes (SMBHs) are thought to exist in almost all massive galaxies \citep{Ferrarese05, Marleau12}.
In the local Universe, the vast majority of these SMBHs are now quiescent, and radiate at many orders of magnitude less than the Eddington luminosity; most notably, this is observed for the Galactic Centre SMBH, Sgr A* \citep{Melia01, Alexander05, Ho09, Genzel10}.

Increased star formation rates (SFRs), along with massive stars, are seen towards the centre of many galaxies \citep{Sarzi04, Walcher06, Schruba11, Kennicutt12, Neumayer12}.
The best studied nucleus containing these features is that of the Milky Way, where massive stars have been observed in a high concentration out to half a parsec from the SMBH \citep[for example,][]{Do13a, Do13b, Lu13}.
It is generally believed that winds from these stars are accreted on to the SMBH through a radiatively inefficient flow, which results in a gas density that is a decreasing power-law function of radius from the SMBH \citep{Quataert04, Cuadra06, Generozov15}.

These massive stars are also the eventual progenitors of core-collapse supernovae (SNe) in galactic nuclei.
Within the sphere of influence of Sgr A* (the region within which the gravitational potential of the SMBH is dominant), evidence for at least one supernova event has been observed in the supernova remnant (SNR) Sgr A East.
This SNR appears to be engulfing Sgr A* with a radius of several $\mathrm{pc}$, and its age has been estimated to be $\sim 10^4~\mathrm{yr}$ \citep{Maeda02, Herrnstein05, Lee06, Tsuboi09}.
Recently, \textit{XMM-Newton} observations over a larger scale in the Galactic Centre have been interpreted to suggest the presence of a second SNR, some $20~\mathrm{pc}$ across \citep{Ponti15}.

In addition, a number of stellar remnants have been detected in the Sgr A* sphere of influence, pointing again towards supernova explosions in the past of this nucleus. 
The Cannonball neutron star (CXOGC J174545.5-285829) has been proposed as originating from the supernova event that created Sgr A East, and is currently close to the edge of the SNR ejecta shell \citep{Park05}; tracing its motion back to the centre of the shell suggests an age of $9000~\mathrm{yr}$ \citep{Zhao13}.
Recently, a magnetar (SGR J1745--2900) was discovered at $\lesssim 2~\mathrm{pc}$ from Sgr A*, and this has been postulated as being associated with the possible $20~\mathrm{pc}$ SNR \citep{Ponti15}.
The presence of stellar remnants within the sphere of influence has also been confirmed from the observation of at least four X-ray binaries (XRBs) within $1~\mathrm{pc}$ of the SMBH, though whether they are high- or low-mass XRBs is uncertain \citep{Muno05}.

Unlike those in the solar neighbourhood, the SNRs in galactic nuclei evolve in an exceptional gaseous environment that is dominated by a radiatively inefficient accretion flow around the SMBH.
The importance of understanding how SNRs evolve in these environments is highlighted, for example, in recent work by \cite{Gaggero15} modelling the Galactic Centre $\gamma$-ray excess using a high supernova rate in the region.
In a previous paper \citep[][hereafter, Paper I]{Rimoldi15}, we developed a numerical shock solver to predict the evolution of SNRs in quiescent galactic nuclei, and showed how their age, size and shape are influenced by the accretion flow properties.
Leveraging the results from that work, here we propose the use of X-ray emission from young SNRs to study the close environment of quiescent SMBHs.

In searches for low luminosity SMBHs, X-rays are often used to avoid extra-nuclear contaminants that affect optical emission \citep[for a recent example, see][]{Miller15}; however, these X-ray searches are not without their own contaminants, and the emission from XRBs is regularly discussed as a prominent contribution.
Yet, if we could observe our Galactic Centre from a more distant perspective, SNR emission would in fact be the dominant contaminant, as the Sgr A East SNR is brighter than both Sgr A* and individually detected XRBs in the sphere of influence.

We are, therefore, interested in how the lifetimes of SNRs in other galactic nuclei can be used to estimate the contribution of SNRs to the nuclear X-ray emission.
Most importantly for us, a dominant contribution from SNRs may allow us to use the detected X-ray emission to constrain the gas properties and SFR.
This information may ultimately help us to understand whether there is a connection between the gas component, the young stellar population and the SMBH that is universal, as well as the nature of this relationship.
More generally, this would be an important step forward in the understanding of the interplay between SMBHs and their host galaxies.

In this paper, we assume that massive star and gas distributions are self-similar in the sphere of influence of quiescent SMBHs, of which our Galactic Centre provides an observational basis.
The universality of this model does not differentiate between SMBH environments by galactic morphology.
Therefore, a discussion of core-collapse SNe in elliptical galaxies may, at first, appear at odds with the current picture of ellipticals.
The morphology of a galaxy is typically a very decisive factor regarding which type of SNe are seen on a galactic scale, and in elliptical galaxies, observational identifications of core-collapse SNe are very rare.
They cannot be ruled out completely, however, as shown by the observation of a probable stripped core-collapse (Type Ib) supernova SN~2005cz in the outskirts of the elliptical galaxy NGC~4589, which appears to have undergone some recent star formation due to a merger \citep{Zhang08, Kawabata10}.

Although most of the volume of elliptical galaxies is devoid of star formation, in the vicinity of the SMBH, star formation may still be present within sufficiently cooled accretion flow of stellar winds (analogous to the possible \textit{in situ} origin of the young stars in the Galactic Centre).
Surveys of the nuclear regions of local elliptical galaxies suggest an inverse correlation between their nuclear activity and the presence of sufficiently cooled interstellar material near the SMBH \citep{Zhang08}.
In the case of active nuclei, the central engine may prevent the cooling of gas, and, in turn, star formation \citep{Werner14}.
Therefore, it is worth reiterating that our results are only in the context of quiescent nuclei.

Justifications and details of our self-regulating, `autarkic' model are given in Section~\ref{sec:autarkic_model}, where we present our framework for quiescent SMBH environments (see also Paper I).
In this context, we then predict the total number of SNRs expected in that region at any given time (Section~\ref{sec:total_numbers}) and their total X-ray luminosity, of which we also assess the detectability (Section~\ref{sec:x-ray}).
Finally, we derive the associated SFRs (Section~\ref{sec:SFR}).
Further elaboration on our findings, and our conclusions, can be found in Section~\ref{sec:conclusion}.

\section{Galactic nuclear environments} \label{sec:autarkic_model}
The spheres of influence of quiescent SMBHs have not experienced major continuous inflows of gas for at least the last $10^7 \sim 10^8~\mathrm{yr}$, roughly the estimated duty cycle of an active galactic nucleus \citep{Shankar09}.
During this time, the SMBH mass and its sphere of influence have not appreciably grown in size, and the life cycles of a few to many generations of massive stars have passed.

After most of the original accretion disc has been consumed, the SMBH starts accreting from the winds of massive stars at a very sub-Eddington level.
The resulting gaseous environment takes the form of an almost spherical, steady-state and radiatively inefficient flow, at least up to a substantial fraction of the sphere of influence \citep{Quataert04, Cuadra06}.

Therefore, massive star and gas properties reflect the current and local environmental conditions within the sphere of influence.
In particular, they have had time to create a steady-state system where massive stars are born from the gas in the local accretion flow and give it back in form of winds and SNe.
Since, from these components, our model describes quiescent nuclei as closed, self-regulating systems, we call this an `autarkic' model.

As a consequence of this autarkic behaviour, massive star and gas distributions should trace each other and their profile be universal among quiescent spheres of influence, with the total number of stars and the accretion rate proportional to the mass of the SMBH.
We therefore expect the same properties, regardless of the global galaxy morphology and assembly history of the nucleus, which should instead be imprinted in the low-mass stellar component of the nucleus.

Due to our vantage point, we have some knowledge of the gas and star distributions in the Galactic Centre.
Practically, we can therefore use those observations (Section \ref{sec:model_MW}) to quantitatively develop a general description of quiescent galactic nuclei (Section \ref{sec:model_nuclei}), extending a method first proposed in Paper I.

\subsection{Galactic Centre observations} \label{sec:model_MW}
The archetypal quiescent galactic nucleus for this work is our Galactic Centre.
The SMBH mass ($\Mbh$) of Sgr A* is $4.3 \times 10^6~\Msun$, resulting in a sphere of influence (hereafter SOI) a few parsecs in radius, within which some $M_* \approx 2 \Mbh \approx 10^7~\Msun$ worth of stars reside \citep{Schodel02, Schodel03, Ghez03, Eisenhauer05, Ghez08, Gillessen09}.

The number density distribution of massive stars in the sphere of influence of Sgr A* appears to follow the form of a two-part power law, broken at a radius defined here as $R_\mathrm{b}$ 
\citep{Buchholz09, Do13a}:
\begin{equation}
n_\mathrm{cc}(R) = \kappa_\mathrm{n} \times
\begin{dcases*}
\left(\frac{R}{R_\mathrm{b}}\right)^{-2} & $R \leq R_\mathrm{b}$ \\
\left(\frac{R}{R_\mathrm{b}}\right)^{-4.5} & $R > R_\mathrm{b}$ ,
\end{dcases*}
\label{eq:massive_star_density}
\end{equation}
for some constant $\kappa_\mathrm{n}$, that will be constrained in Section~\ref{sec:total_numbers}.
The steepness of the gradient outside $R_\mathrm{b}$ is more uncertain, due to the low number of stars at this distance; however, for the same reason, the value of the outer gradient does not have a substantial influence on our results.

The gaseous environment in the SOI is dominated by the accretion flow.
The measured density at approximately the scale of the Bondi radius ($\sim 0.04~\mathrm{pc}$) is $\sim 130~\mathrm{cm}^{-3}$, where the mass flow has an Eddington ratio of $\dot{M}/\dot{M}_{\rm Edd} \approx 10^{-5}$ \citep{Baganoff03, Wang13}.
We take this radius, hereafter referred to as $R_0$, as a reference point for the density.
A break in the gas density is expected at $R_\mathrm{b} \approx 0.4~\mathrm{pc}$ where the density of high-mass stars drops off \citep{Quataert04, Cuadra06}.
Within $R_\mathrm{b}$, the density gradient depends on the mode of energy transport.
In standard advection-dominated accretion flows, the inner power law follows $\omega_\mathrm{in} = 3/2$ \citep{Narayan95a, Narayan95b}.
For convection-dominated flows \citep{Quataert00, Ball01} or those with substantial outflows, as in the adiabatic inflow--outflow solution \citep[ADIOS;][]{Blandford99, Begelman12}, the inner gradient is shallower at $\omega_\mathrm{in} = 1/2$.
Although more recent observations tend to favour a density gradient of $R^{-\omega_\mathrm{in}}$ with $\omega_\mathrm{in} = 1/2$ \citep{Wang13}, we also explore the whole possible range $\omega_\mathrm{in} \in \left\{1/2,\; 1,\; 3/2\right\}$.
Outside $R_\mathrm{b}$, instead, we follow results from simulations and we take $R^{-3}$ \citep{Quataert04, Cuadra06}.

Finally, it is now well established that a molecular torus exists around Sgr A*, which extends from just inside the SOI ($\sim 2~\mathrm{pc}$) to about $5~\mathrm{pc}$ from the SMBH \citep{Jackson93, Christopher05, Liu13}.
The torus has a wedge-like profile, where the inner edge is narrower ($\sim 0.4~\mathrm{pc}$ thick) than the outer edge ($\sim 2~\mathrm{pc}$), and contains molecular hydrogen with a density of $n_{\mathrm{H}_2} \approx 10^4~\mathrm{cm}^{-3}$.

\subsection{Quiescent galactic nuclei as autarkic systems} \label{sec:model_nuclei}
We now consider environments of other quiescent galactic nuclei, and we show how their properties can be scaled with the mass of the SMBH (see also Paper I).

The particular region we are most interested in is the SMBH sphere of influence, which contains a total mass in star of $M_* \approx 2 \Mbh$, and the size of which can be estimated as a function of SMBH mass, using the $M$--$\sigma$ relation \citep{Ferrarese00, Gebhardt00}:
\begin{equation} \label{eq:soinosigma}
R_\mathrm{SOI} \approx 2.7 \left(\frac{\Mbh}{4.3 \times 10^6~\Msun} \right)^{7/15}~\mathrm{pc} .
\end{equation}
Our reference value for the Milky Way SOI radius is $R_\mathrm{SOI,MW} = 2.7~\mathrm{pc}$.
In the self-similar spirit of our model, we will also scale the break and density reference radii ($R_{\rm b}$ and $R_0$) proportionally with the SOI size.
As for the Milky Way, we associate $R_0$ with the Bondi radius.
This scaling with $R_\mathrm{SOI}$ therefore implies that the temperature of the gas is proportional to $\sigma^2$.
We will comment on some implications of this later in this section.

Within the sphere of influence, the number density distribution of massive stars has the form of equation (\ref{eq:massive_star_density}), and the total number of these stars is $N_{\rm cc} \propto M_* \propto \Mbh$.
This will be quantified in Section~\ref{sec:total_numbers}, where we will predict the associated steady-state supernova rate and compare with observations. 

The gas density profile around the black hole is universally set by accretion physics for a radiatively inefficient flow, and it is described in the previous section.
The number density, $n$, however, should be estimated through the continuity equation,
\begin{equation}
\dot{M} \approx 4 \uppi R^2 \, m_{\rm p} n(R) \, v_\mathrm{K}(R) ,
\end{equation}
where the radial velocity in a geometrically thick accretion flow is approximately the Keplerian value $v_\mathrm{K}$.
Since the accretion rate $\dot{M}$ is powered by stellar winds, it increases with the stellar number and therefore with the black hole mass in a proportional fashion, $\dot{M} \propto \Mbh$.
This implies that such self-similar quiescent SMBHs emit at the same Eddington ratio.
It follows that the number density in terms of the Milky Way value at the radius $R_0$ is
\begin{equation} \label{eq:gas_number_density}
n(R_0) \approx 130 \left( \frac{\Mbh}{4.3 \times 10^6 \Msun} \right)^{1/2} 
\left(\frac{R_\mathrm{SOI}}{R_\mathrm{SOI,MW}}\right)^{-3/2}~\mathrm{cm^{-3}}.
\end{equation}
Equation (\ref{eq:gas_number_density}) allows us to express the density distributions around other quiescent SMBHs purely as a function of their mass.

Although other kinds of scaling are possible, this simple scaling with $\Mbh$ is consistent with recent, more in depth treatments of quiescent SMBH circumnuclear media \citep[see][where their stagnation radius is comparable to the Bondi radius]{Generozov15}.
Moreover, our scaling of $R_0$ with $\Mbh$ (such that $R_0 \ll R_\mathrm{b}$) is compatible with the results in \cite{Generozov15} in the high-heating limit, which corresponds to continuous star formation in their work \citep[which is also assumed here based on observational evidence in the Milky Way;][]{Figer04, Figer09, Pfuhl11}.

In all these galactic nuclei, the density is expected to flatten from the $R^{-3}$ gradient around the scale of the SOI.
In this paper, we more carefully model the ambient density near and beyond the SOI with respect to Paper~I.
We considered a few possible variations for the way the density levels off: a floor of $1~\mathrm{cm}^{-3}$ (irrespective of radius), a fixed value of $1~\mathrm{cm}^{-3}$ beyond $R_\mathrm{SOI}$, and a fixed value of $n (R > R_\mathrm{SOI}) = n (R_\mathrm{SOI})$.
Regardless of the choice, we found the variations in our final results (such as the variation in the predicted temperatures of Section \ref{sec:spectral_properties}) were minimal.
In the remaining work, we impose a floor in the density at the value $n = 1~\mathrm{cm}^{-3}$.

In this paper, we additionally embed a molecular torus within the power-law ambient medium.
The torus is taken to begin at $R_\mathrm{SOI}$ and extend to $5 \left( R_\mathrm{SOI} / R_\mathrm{SOI,MW}~\mathrm{pc} \right)$ from the SMBH, with the inner and outer thicknesses described as above (also scaled by $R_\mathrm{SOI} / R_\mathrm{SOI,MW}$).
The density within the torus is taken to be $2 \times 10^{4}~m_\mathrm{p}~\mathrm{cm}^{-3}$ independent of the SMBH mass, as it is a property of the molecular cloud.

\section{SNR dynamical evolution}
For our purposes, we need to trace the evolution of SNRs (including their morphology and shock velocity) in the ambient medium of galactic nuclei explained in Section~\ref{sec:autarkic_model}.
To this end, we use the method developed in Paper I, where the reader can find a detailed description. 

In short, this method exploits the Kompaneets approximation to follow the evolution of a strong shock from a SNR in an axisymmetric configuration of density.
In this paper, we explode the SNRs at different distances along the axis of symmetry of the molecular torus.
Along with the power-law background, this preserves the axisymmetry of the problem that was originally exploited in the design of our code.

Once the shock decelerates, the temperature of the shocked gas becomes sufficiently low for line cooling to efficiently radiate energy from the SNR.
Prior to this stage, the SNR is deemed `adiabatic', as the energy lost is a very small fraction of the total energy in the shocked gas.
We define the end of the adiabatic stage to occur when the SNR has succumbed to one of two outcomes: either $\geq 50$ per cent of the SNR, measured by ejecta mass fraction, has reached this radiative stage ($T \lesssim 10^6~\mathrm{K}$; $v \lesssim 300~\mathrm{km\,s}^{-1}$), or $\geq 50$ per cent has been sheared apart from decelerating enough that the velocity is comparable to the local Keplerian velocity around the SMBH.
If deceleration is not appreciable, then the SNR shearing happens at a radius
\begin{equation} \label{eq:shearing_radius}
R_\mathrm{sh} = 1.9 \times 10^{-4} \left(\frac{\Mbh}{4.3 \times 10^6 \Msun}\right) \left(\frac{v_\mathrm{init}}{10^4~\mathrm{km \, s}^{-1}}\right)^{-2}~\mathrm{pc },
\end{equation}
where we assumed an ejection velocity of $10^4~\mathrm{km \, s}^{-1}$.
In all our calculations, $R_\mathrm{sh}$ is the minimum explosion radius at which a SNR can survive.

We have found that including a molecular torus in our simulations does not have a large effect on the SNR dynamics or morphology, as the shock front effectively diffracts around the barrier and continues its outward motion after self-intersecting on the far side.
Therefore, we do not expect tori of the dimensions considered here to confine or strongly shape the SNR once it has expanded to the scale of the sphere of influence.

\subsection{X-ray emitting lifetime} \label{sec:lifetimes}
Improving on Paper I, the more careful modelling of the environment just outside the SOI allows us to more robustly quantify the adiabatic lifetime for SNRs that expand beyond the SOI, and survive through to the radiative stage.
We find that, regardless of the specific choice in the way the density flattens (Section~\ref{sec:model_nuclei}), the adiabatic stage ends after a similar time, around $2 \times 10^4~\mathrm{yr}$.

\begin{figure}
\centering
\includegraphics[width=1.08\columnwidth]{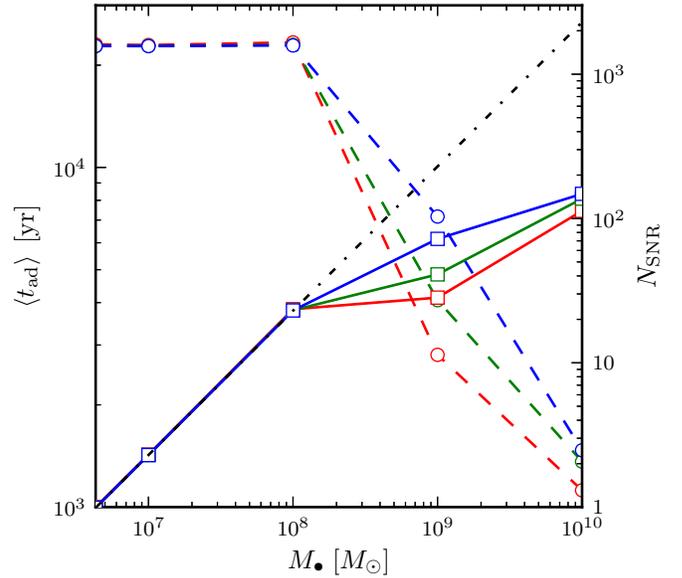}
\caption{Circles with dashed lines (left-hand axis) show the mean adiabatic lifetime of SNRs as a function of $\Mbh$ and $\omega_\mathrm{in}$, measured by shearing a total of $M_\mathrm{ej}/2$.
By $\sim 2 \times 10^4~\mathrm{yr}$, radiative losses become significant in all cases; if the shearing condition $v_{\rm sh} < v_{\rm K}$ has not yet been met by this time, the adiabatic stage ends due to the radiative transition (indicated by a dotted black line at low $\Mbh$.
The squares with solid lines (right-hand axis) show the expected number of adiabatic SNRs in galactic nuclei as calculated in equation (\ref{eq:expected_number}), with a scaling based on the observation of one core-collapse SNR in the Milky Way.
In each case, the red (lowermost) line corresponds to an inner gas gradient $\omega_\mathrm{in} = 1/2$, the green line is for $\omega_\mathrm{in} = 1$ and the blue (uppermost) line is for $\omega_\mathrm{in} = 3/2$.
The reduction in lifetime by SMBH shearing is compared with the value expected if SNRs ended their adiabatic stage from radiative losses alone ($\sim 2 \times 10^4~\mathrm{yr}$), shown as a black dot--dashed line.}
\label{fig:lifetimes_and_numbers}
\end{figure}

More generally, we calculate the adiabatic lifetime $t_{\rm ad}(R)$ of a SNR as a function of distance within the SOI, regardless of its fate (whether sheared or not). 
We then compute, for a given black hole mass, the  mean adiabatic lifetime $\left < t_{\rm ad} \right>$  by weighting $t_{\rm ad}(R)$ by the number density of massive stars at that location (equation~\ref{eq:massive_star_density}; for more detail see section~5.3 in Paper I).
The result is shown in Fig.~\ref{fig:lifetimes_and_numbers}, with the circles and dashed lines (left-hand axis).
The three inner density gradients $\omega_\mathrm{in} \in \left\{1/2,\; 1,\; 3/2\right\}$ are shown, where the red line, $\omega_\mathrm{in} = 1/2$, corresponds to that favoured by observations in the Milky Way.
For $\Mbh < 10^{8}~\Msun$, the average SNR ends its life evolving through the radiative phase, while for $\Mbh > 10^{8}~\Msun$ the combination of ambient medium deceleration and black hole tidal forces disperse the SNR before the radiative stage, shortening the duration of its X-ray emitting phase.
The suppression of adiabatic lifetime increases with $\Mbh$ and at $\Mbh = 10^9~\Msun$ is an order of magnitude smaller for $\omega_\mathrm{in} = 1/2$.

\section{Number of adiabatic remnants in a snapshot observation} \label{sec:total_numbers}
Using results for the lifetimes of SNRs as a function of distance within the SOI, we proceed to calculate the average number ($N_\mathrm{SNR}$) of core-collapse SNRs that could be observed at any given time in X-ray, within the SOI of an SMBH.
Knowledge of $N_\mathrm{SNR}$ will allow us to later determine the expected contribution of young SNRs to the X-ray emission near the SMBH.

We assume here that the accretion flow refilling time-scale from stellar winds is similar to, or shorter than, the supernova rate, such that, on average, previous SNRs do not significantly effect the gaseous environment of subsequent SNRs.
This appears to be the case for the Milky Way, where simulations show that a quasi-steady-state gas distribution matching the one we assume here is reached over a time-scale ($2 \times 10^3~\mathrm{yr}$), which is an order of magnitude shorter than the expected supernova rate of one per $\sim 10^{4}~\mathrm{yr}$ \citep{Cuadra06}.
In more massive nuclei, we expect the time between SNe to shorten proportionally to the mass input from stellar winds, sustaining the competing effects of supernova sweeping and wind refilling.

In a steady state case, the massive stars are replenished by star formation at the same rate as they explode as SNe, and their number at any location is independent of time.
In this case, $\ud N (M>8) = n_\mathrm{cc}(R) R^2 \, \ud R$ in any spherical shell of distance $R$ from the SMBH.\footnote{We assume that the location of the SMBH coincides with the centre of the stellar distribution.}
The number of `adiabatic' remnants, $N_\mathrm{SNR}$, expected at any time in the SOI is thus
\begin{equation}
N_\mathrm{SNR} = 4 \uppi \int_{R_\mathrm{sh}}^{R_\mathrm{SOI}} { \frac{n_\mathrm{cc}(R) \, t_\mathrm{ad}(R)}{\langle t_*(M\,{>}\,8)\rangle } R^2 \, \ud R } ,  
\label{eq:expected_number}
\end{equation}
where $\langle t_*(M\,{>}\,8)\rangle $ is the stellar lifetime $t_*(M) \approx 10^{10} \left( M / \Msun \right)^{-2.5}~\mathrm{yr}$, weighted over the stellar initial mass function (IMF), $\varphi(M)$, for $M > 8\,\Msun$.
As for the stellar and gas distributions, we take a universal current IMF in black hole SOIs. 

To solve generally for $N_\mathrm{SNR}$, the total number of massive stars, $N_\mathrm{cc} =  4 \uppi \int_{R_\mathrm{sh}}^{R_\mathrm{SOI}} n_\mathrm{cc} R^2 \ud R $, as a function of $\Mbh$ is required.
This will allow us to determine $\kappa_\mathrm{n}(\Mbh)$ in the definition of $n_\mathrm{cc}$ (equation \ref{eq:massive_star_density}), which in turn is to be used in equation \ref{eq:expected_number}.
For $\Mbh = 4.3 \times 10^6\,\Msun$, equation (\ref{eq:expected_number}) is simplified by the fact that our simulations, as summarized in Fig.~\ref{fig:lifetimes_and_numbers}, show $t_\mathrm{ad}(R)$ to be constant ($2 \times 10^4~\mathrm{yr}$) over $R$ within the SOI, since there is no explosion location where SNRs are destroyed by shearing.
We can therefore divide through by $t_\mathrm{ad}(R)$ and write equation (\ref{eq:expected_number}) as a supernova rate:
\begin{equation}
\mathcal{R}_\mathrm{SN} = \frac{N_\mathrm{SNR}}{t_\mathrm{ad}(R)} = \frac{N_\mathrm{cc} }{\langle t_*(M\,{>}\,8) \rangle}.
\label{eq:supernova_rate}
\end{equation}
From these equalities, we estimate $N_\mathrm{cc}$, using observations to set a value for $\langle t_*(M\,{>}\,8) \rangle$ and $N_\mathrm{SNR}$, as explained below.

For $\langle t_*(M\,{>}\,8) \rangle$, the mass distribution of core-collapse progenitors is needed.
\cite{Pfuhl11} find that the long-term star formation in the Milky Way nuclear star cluster is best fit by an approximately Chabrier/Kroupa IMF,
\begin{align} \label{eq:imf}
\varphi(M) &= \frac{\ud N}{\ud M} \propto M^{-\alpha} , \\
\alpha     &=
\begin{dcases*}
1.3, & $0.1 \, \Msun \leq M \leq 0.5 \, \Msun$ \\ \nonumber
2.3, & $0.5 \, \Msun < M$. \\
\end{dcases*}
\end{align}
With this IMF, the mean stellar lifetime for core-collapse progenitors is 
\begin{align}
\langle t_*(M>8)\rangle &= \frac{10^{10} \int_{8 \Msun}^{50 \Msun} \left( M/\Msun \right) ^{-(2.5 + \alpha)} \ud M }{\int_{8 \Msun}^{50 \Msun} \left( M/\Msun \right)^{-\alpha} \ud M }~\mathrm{yr} \nonumber \\
                        &= 2 \times 10^7~\mathrm{yr} ,
\label{eq:t_star}
\end{align}
where stars above $\sim 50~\Msun$ are taken to form black holes directly without a corresponding supernova \citep{Fryer99, Yungelson08}.\footnote{The numerical result does not change appreciably if the integration limit is $\infty$.}
Note that this mass function does not describe the current stellar content of the Galactic Centre, since the majority of the total stellar mass is in the longer-lived low-mass stars, most of which formed more than $5~\mathrm{Gyr}$ ago \citep{Pfuhl11}.
The present-day mass function is also modified by accumulated stellar remnants. 

As mentioned in Section \ref{sec:model_MW}, there are one or two potential SNRs within the SOI of Sgr A*: the Sgr A East shell and a possible SNR suggested by \cite{Ponti15} in observations of the $\sim 20~\mathrm{pc}$ X-ray emitting lobes.
Sgr A East has been argued to be a $\lesssim 10^4$-year-old Type II SNR that is transitioning to the radiative phase \citep{Maeda02}.
The $20~\mathrm{pc}$ structure may be an SNR of similar age, possibly associated with the $\sim 10^4$-$\mathrm{yr}$-old magnetar SGR~J1745--2900 in the sphere of influence \citep{Ponti15}.
Taking at least one of these two possible SNRs to have been generated by a core-collapse supernova in the SOI, we set $N_\mathrm{SNR} = 1$.
Through equation (\ref{eq:supernova_rate}), we then derive $N_\mathrm{cc} \approx 1000$ for the Milky Way.

There is, however, evidence to suggest that the IMF of, at least, the recently formed stellar disc(s) is more top-heavy \citep[$\alpha \approx 0.45$;][]{Paumard06, Bartko10}.
Therefore, we also consider the effect of using $\alpha = 0.45$ in equation (\ref{eq:imf}).
This reduces the mean stellar lifetime to $\langle t_*(M>8) \rangle = 9 \times 10^6~\mathrm{yr}$, and, therefore, $N_\mathrm{cc}$ is reduced to $\sim 500$.
These values of $N_\mathrm{cc}$ are slightly higher than the number of sufficiently massive stars 
found in recent censuses of the inner half parsec \citep[around a few hundred;][]{Do13b}, though some discrepancy may be expected if current $K$-band spectroscopic limits restrict these observations to very early-type stars and the innermost region \citep{Lu13}.

Since $N_\mathrm{cc}$ is proportional to the total stellar mass and $M_* \propto \Mbh$, the scaling with the mass of the SMBH is simply
\begin{equation}
N_\mathrm{cc} \approx 10^3 \left (\frac{\Mbh}{4.3 \times 10^6 \Msun} \right).
\label{eq:number_cc}
\end{equation}
Equation~(\ref{eq:expected_number}) can now be solved generally for quiescent nuclei as a function of $\Mbh$, and our result is shown in the solid lines (right-hand axis) of Fig.~\ref{fig:lifetimes_and_numbers}.
The number of observed SNRs at any given time grows with $\Mbh$ from $N_\mathrm{SNR} = 1$ for $\Mbh = 4.3 \times 10^6~\Msun$ to around $10^{2}$ for $\Mbh = 10^{10}~\Msun$.
This trend is the result of two competing effects: as $\Mbh$ increases, SNR lifetimes become shorter but the total number of massive stars increases (equation~\ref{eq:number_cc}).
Since the latter grows faster, the net behaviour is a positive gradient.

The dot--dashed black line in Fig.~\ref{fig:lifetimes_and_numbers} shows a comparison with the case if SNR lifetimes were not ended by shearing, but instead continued through a final radiative phase ($\gtrsim 10^4~\mathrm{yr}$), as is typical in a constant ISM.
In this case, $N_\mathrm{SNR} \propto \Mbh$, while our results (solid lines) show a sublinear growth.
The reduction is most prominent at the highest masses, where the mean X-ray lifetime of a SNR is several times smaller than $10^4~\mathrm{yr}$.
A spread in the expected number of remnants can be seen to be dependent on the choice of inner gas gradient around $\Mbh = 10^9~\Msun$, where the red and blue lines differ by a factor of 3.
We will show later that most of local galaxies suitable for X-ray observations have SMBHs around that mass (Section~\ref{sec:detectability}).
Therefore, our result suggests that, in principle, it may be possible to use these nuclei to probe the inner accretion flow.
We will elaborate on this point later, when we discuss our results on the expected total luminosities (Sections~\ref{sec:detectability} and \ref{sec:conclusion}).

In the remainder of this paper, however, we will only present results for  $\omega_\mathrm{in} = 1/2$, as this is the most favoured value from recent observations of the Galactic Centre \citep{Wang13}. Also, this gives a conservative lower limit for our predictions, and, as will be apparent later, our luminosity estimates are more uncertain than the difference between results from alternative gas density profiles.

\section{X-ray luminosity from SNRs in the sphere of influence} \label{sec:x-ray}
Simultaneous SNRs in their adiabatic phase (Fig.~\ref{fig:lifetimes_and_numbers}) should contribute to the X-ray emission from quiescent nuclei.
Here, we aim to quantify their total soft and hard X-ray luminosity, compare it to other sources and assess its detectability. 
In the X-ray band, the SNR dominant emission mechanism is bremsstrahlung radiation, unless the SNR is sweeping into a very rarefied environment \citep{Vink12}.

Therefore, to make an estimate of their relative brightness in the hard and soft band, we first make predictions of the gas temperature behind the SNR shocks.
We calculate this temperature as a function of time as well as the most probable age of a SNR in a single observation, as very young SNRs are hotter than old ones.

\subsection{SNR spectral properties} \label{sec:spectral_properties}
We characterize the temperature of SNRs in the sphere of influence of an SMBH by analysing data from the simulations outlined in Paper I.
The challenge is that the medium is not uniform and different parts of the remnant hold different temperatures and luminosities, and these quantities change with time.
Since we want to characterize the emission in a snapshot observation, we need a measure of the temperature that most contributes to the SNR 
spectrum, at its most probable age.
We proceed as follows.

We first determine the temperature behind each point along the shock using the shock front velocity, $v_\mathrm{s}$, via
\begin{equation}
T' = \frac{3 m_\mathrm{u} \mu}{16 k_\mathrm{B}} v_\mathrm{s}^2 ,
\end{equation}
for an adiabatic exponent of $\gamma = 5/3$, and where $m_\mathrm{u}$ is the atomic mass unit, $\mu$ is the mean molecular mass, and $k_\mathrm{B}$ is Boltzmann's constant.

At a given moment in time, the SNR mean temperature is found by weighting the temperature behind the shock, $T'$, at each point along the shock front, by the rate of radiative cooling over the line-dominated (low temperature) and bremsstrahlung-dominated (high temperature) regimes,
\begin{equation} \label{eq:cooling_function}
\Lambda_\mathrm{li,br}(\rho', T') \propto
\begin{dcases*}
\left(\rho'\right)^2 \left(T'\right)^{-1}, & $T \leq 3 \times 10^7~\mathrm{K}$ \\
\left(\rho'\right)^2 \left(T'\right)^{1/2}, & $T > 3 \times 10^7~\mathrm{K}$  ,
\end{dcases*}
\end{equation}
where $\rho'$ is the density behind the shock.
For a strong shock, the postshock density is found simply from the compression ratio $\rho'/\rho = 4$.
Next, $T'$ is weighted by the surface area, `$A$', of each section of the SNR, which is the conical frustum (excluding circular caps) obtained by rotating the small cross-sectional segments of the shock front at that position about the axis of symmetry.
The spatial mean of $T'$ is computed along the entire evolution of the SNR.
\begin{figure}
\centering
\includegraphics[width=\columnwidth]{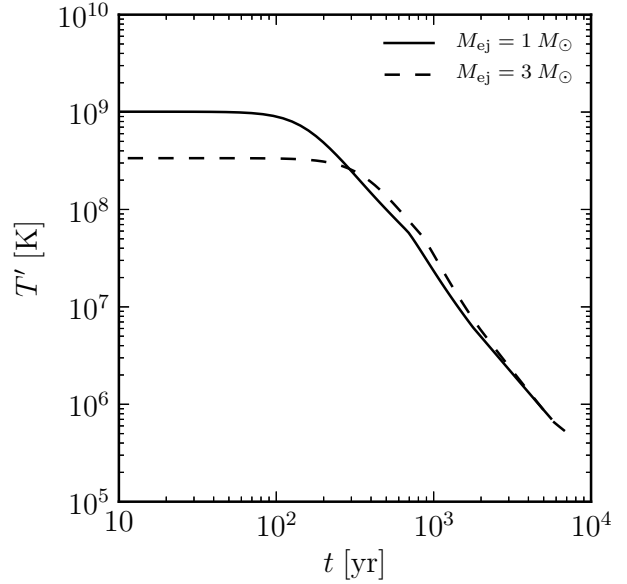}
\caption{Temperature of an SNR, averaged by emissivity over the SNR surface as described in Section~\ref{sec:spectral_properties}.
The SNR shown here explodes at $10~\mathrm{pc}$ from an SMBH of mass $10^{9}~\Msun$.
The solid line represents an ejecta mass behind the shock of $1~\Msun$, while the dashed line shows a mass of $3~\Msun$.
In each case, the adiabatic evolution stops when more than $50$ per cent of this mass is tidally sheared (which defines the last time point on this figure).}
\label{fig:temperature_example}
\end{figure}

Two examples of this temperature evolution are plotted in Fig.~\ref{fig:temperature_example}, both at an explosion distance of $10~\mathrm{pc}$ away from an SMBH of mass $\Mbh = 10^{9}~\Msun$.
They clearly show how initially a supernova may be tens of $\mathrm{keV}$ hot, while thousands of years later its temperature can be well below $1~\mathrm{keV}$.
The SNR depicted with the solid line has an ejecta mass of $1~\Msun$, while the dashed line shows the case for an ejecta mass of $3~\Msun$, to investigate variation in ejecta mass behind the shock front.

The shock velocity at the start of the ejecta-dominated stage in our simulations is determined by depositing the $\sim 10^{51}~\mathrm{erg}$ of explosion energy as kinetic energy into the given ejecta mass.
Therefore, the higher mass of ejecta has a lower initial velocity, but takes longer to decelerate due to the need to sweep up more material before reaching the adiabatic stage.
We ran simulations over all the $\omega = 1/2$ initial conditions with $M_\mathrm{ej} = 3~\Msun$ for comparison.
We find that the late-time evolution of the SNR is relatively indifferent to the ejecta mass, resulting in a negligible difference in the SNR lifetimes compared to those with $M_\mathrm{ej} = 1~\Msun$.

Additionally, we weight the spatial mean by the time spent at that temperature ($\ud t$, at the resolution of the simulation snapshots), giving
\begin{equation} \label{eq:temperature_weighted_at_r}
\langle T'(R) \rangle = \frac{ \iint T' \, \Lambda_\mathrm{li,br}(\rho', T') \, \ud A \, \ud t}{\iint \Lambda_\mathrm{li,br}(\rho', T') \, \ud A \, \ud t} .
\end{equation}
This is the expected temperature observed in a single observation of a SNR, exploding at a given radius $R$ from the SMBH.

\begin{figure}
\centering
\includegraphics[width=1.08\columnwidth]{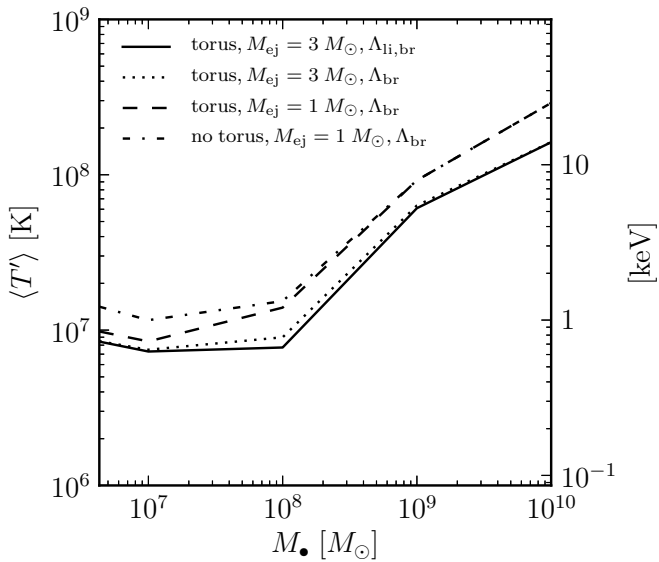}
\caption{Mean temperature (left-hand axis) and corresponding energy (right-hand axis) for X-ray emitting SNRs exploding in the sphere of influence as a function of SMBH mass.
Weighted mean temperatures are calculated using equation (\ref{eq:temperature_weighted_at_r}) at a given explosion distance, and then equation (\ref{eq:temperature_weighted_per_SMBH}) for the entire sphere of influence.
Four cases are compared.
The solid line shows a molecular torus environment as outlined in Section~\ref{sec:autarkic_model} with $M_\mathrm{ej} = 3~\Msun$ and a cooling function ($\Lambda_\mathrm{li,br}$) that transitions between the line cooling ($T^{-1}$) and bremsstrahlung ($T^{1/2}$) relations (equation \ref{eq:cooling_function}).
The remaining three curves are calculated with a purely bremsstrahlung cooling relation ($\Lambda_\mathrm{br}$): the same molecular torus profile again with $M_\mathrm{ej} = 3~\Msun$ (dotted line), a density profile with a molecular torus and $M_\mathrm{ej} = 1~\Msun$ (dashed line) and a simple power-law density profile with no torus (dot--dashed line) and $M_\mathrm{ej} = 1~\Msun$.}
\label{fig:temperatures}
\end{figure}

Finally, we calculate the expected temperature of young SNRs in a given galactic nucleus by weighting $\langle T'(R) \rangle$ by the likelihood of a core-collapse supernova at each location, which is proportional to the number density of massive stars, $n_\mathrm{cc}$:
\begin{equation} \label{eq:temperature_weighted_per_SMBH}
\langle T' \rangle = \frac{\int_\mathrm{R_\mathrm{sh}}^{R_\mathrm{SOI}} \langle T'(R) \rangle \, n_\mathrm{cc}(R) \, R^2 \, \ud R}{\int_\mathrm{R_\mathrm{sh}}^{R_\mathrm{SOI}} n_\mathrm{cc}(R) \, R^2 \, \ud R} .
\end{equation}
This gives the expected temperature per galactic nucleus, which we plot in Fig.~\ref{fig:temperatures}.

In Fig.~\ref{fig:temperatures}, we consider the effects of not only adding the torus and varying the ejecta mass, but also simplifying the cooling function used.
The solid and dotted lines of Fig.~\ref{fig:temperatures} show that replacing the cooling function by a purely bremsstrahlung form, $\Lambda_\mathrm{br}(\rho', T') \propto (\rho')^2 (T')^{1/2}$, produces only a very minor difference in the result (the density in the ambient medium when the SNR has appreciably decelerated also tends to be low, which reduces the low temperature emissivity).
Therefore, for simplicity in the rest of this work, we perform our calculations with the bremsstrahlung-dominated function, $\Lambda_\mathrm{br}(\rho',T')$.

There is a clear trend in Fig.~\ref{fig:temperatures} from low to high $\langle T' \rangle$ as the SMBH mass increases.
This is due to the shortening of SNR X-ray lifetimes with increasing SMBH mass discussed in Section~\ref{sec:lifetimes}.
SNRs with shortened lifetimes do not spend a long time as cooler, softer X-ray objects, and therefore the expected temperature of an observed SNR is higher.
On the other hand, for $\Mbh$ closer to that of Sgr A*, the SNRs do evolve through to the radiative stage, and spend much of their adiabatic life in the softer X-ray stage, reducing the overall expected temperature.

This effect suggests that SNRs around more massive SMBHs will have an influence on harder X-ray observations (for observations that extend to these high energies), while SNRs around lower mass SMBHs are more likely to influence soft X-ray observations.
To test the robustness of our results, we again investigate the effect of varying the ejecta mass.
We also test the impact of imposing the molecular torus on the density profile.
It is evident from all the plotted curves that the final result does not strongly depend on either the presence of a torus or the ejecta mass.

As a caveat, we note that the electron temperature important for bremsstrahlung emission behind the shock front is dependent on the degree of energy equipartition with the shocked ionic component.
There is some debate on the degree of equipartition in observed SNR plasmas, in part motivated by the fact that the thermal bremsstrahlung from very young SNRs has not been seen to exceed about $4~\mathrm{keV}$~\citep[for a recent review, see][]{Vink12}.
Therefore, the hotter post-shock temperatures predicted here may be somewhat suppressed when considering the electron temperature relevant for radiative processes.

For comparison with observations, we note that the spectrum of Sgr A East has been described with either a plasma with a $k_\mathrm{B} T \approx 2~\mathrm{keV}$ electron temperature \citep{Maeda02} or a two-temperature, thermal plasma of $1$ and $4~\mathrm{keV}$ \citep{Sakano04}.
The temperature of the bipolar lobes, and possible second SNR, in the Galactic Centre is also fit with a hot component of a comparable value \citep{Ponti15}.
A pervasive X-ray emission at $\sim 1~\mathrm{keV}$ is well-known to exist throughout the Galactic Centre region, which has also been attributed to SNRs \citep{Muno04, Ponti15}.
These temperatures are in good agreement with the mean expected value for a SNR in the Galactic Centre predicted here.

\subsection{SNR X-ray luminosity} \label{sec:x-ray_luminosity}
With a prediction of the total number of SNRs in a sphere of influence as well as their mean temperatures, we can now consider the total integrated X-ray emission from SNRs.
As in the previous section, we assume that each SNR contributes an X-ray luminosity that is unaffected by previous SNRs.

Our simulations focus on the dynamical properties of the SNR shock fronts, which allowed us to determine the post-shock temperature in Section~\ref{sec:spectral_properties}.
However, in the absence of, at least, a detailed model of plasma properties within the SNR volume as well as associated radiative processes to predict the luminosity from first principles, we turn to X-ray observations of young SNRs to guide our estimates for the total luminosities.

There is a large variation in luminosities observed for young core-collapse SNRs \citep{Dwarkadas12}.
Some of this variation can be attributed to different supernova types.
For example, Type IIn SNe are generally brighter in X-rays in the initial $10^3 \sim 10^4~\mathrm{d}$ compared to other core-collapse types;\footnote{This, along with the narrow hydrogen emission lines observed in their spectra, is attributed to interaction with high-density surrounding material.} however, even just within the Type IIn classification, there can be variations of orders of magnitude in the early X-ray luminosity.
For the most common core-collapse SNe, Type IIP, X-ray luminosities for Galactic SNRs have been estimated to start at $\sim 10^{38}~\mathrm{erg\,s}^{-1}$ and decrease up to an order of magnitude within the first $\sim 10^3~\mathrm{d}$ \citep{Dwarkadas12}.

The compilation in \cite{Dwarkadas12} reports $L_\mathrm{X}$ over a range of different bands dependent on the X-ray observatory used.
For a broader set of data, we also consider the \textit{Chandra} ACIS Survey of M33 (ChASeM33), which studied a large number of young SNRs \citep{Long10}.
With a survey threshold of $L_\mathrm{X,\,0.35-2\,keV} \approx 2 \times 10^{34}~\mathrm{erg\,s}^{-1}$ (as well as the fact that M33 is a large, face-on spiral belonging to the Local Group) this provides a large sample of known extragalactic SNRs in soft X-rays.
A total of 137 SNRs and SNR candidates were identified, with a median diameter of $44~\mathrm{pc}$ (comparable to a middle-aged SNR) and inferred luminosities of $2.4 \times 10^{34}\sim 1.2\times10^{37}~\mathrm{erg\,s}^{-1}$ in the soft X-ray band ($0.35$--$2~\mathrm{keV}$), with increasing number density for decreasing $L_\mathrm{X,\,0.35 - 2\,keV}$.
The brightest of these were of intermediate diameter ($15 \sim 40~\mathrm{pc}$) but exhibited localised, enhanced X-ray emission suggestive of interactions with denser material.

There is some variation in the numbers of SNRs at the uppermost values of $L_\mathrm{X,\,0.35-2\,keV}$ in M33 compared to the Magellanic Clouds.
As suggested in \cite{Long10}, this variation may be attributed to small-number statistics, though there may also be some variation due to differing galactic morphological types.
\cite{Long10} note that three well-known, young SNRs in the Milky Way---Cas A, Kepler and Tycho---are all emitting at around a few $10^{36}~\mathrm{erg\,s}^{-1}$ in the $0.35$--$2.0~\mathrm{keV}$ band.
\begin{figure}
\centering
\includegraphics[width=\columnwidth]{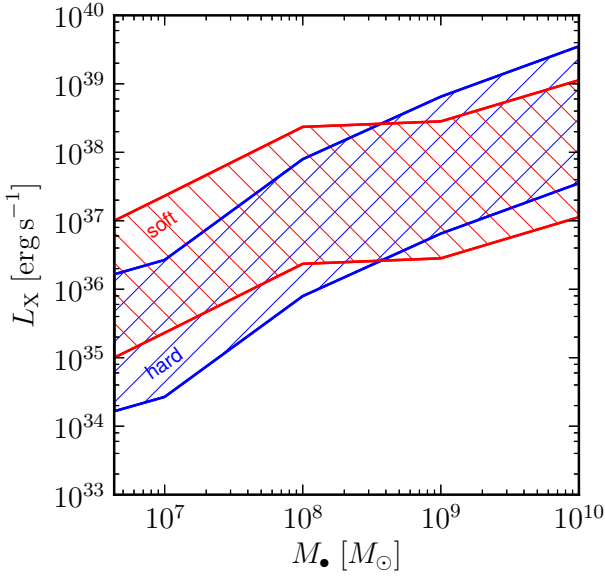}
\caption{X-ray luminosities for SNRs in the soft ($0.35$ to $2~\mathrm{keV}$) and hard ($2$ to $8~\mathrm{keV}$) bands.
The red back-hatched (`\textbackslash') band shows the soft band limits determined from observations of young SNRs.
The blue forward-hatched (`/') band shows the range of $L_\mathrm{X}$ expected in the hard band; as for Fig.~\ref{fig:luminosity_hard}, this emission is determined from the soft band luminosities scaled using a thermal bremsstrahlung spectrum with the temperature found in Section~\ref{sec:spectral_properties}.}
\label{fig:luminosity_both}
\end{figure}
\begin{figure}
\centering
\includegraphics[width=\columnwidth]{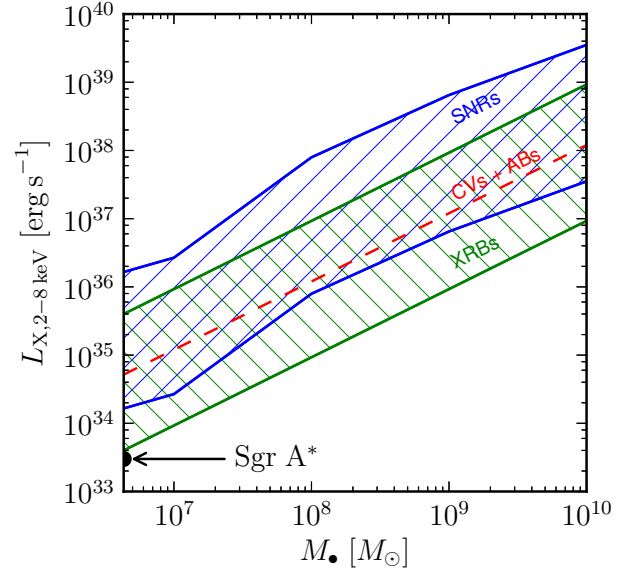}
\caption{X-ray luminosities in the $2$--$8~\mathrm{keV}$ band for SNRs and XRBs in quiescent SOIs.
The blue forward-hatched (`/') band shows the limits of $L_\mathrm{X}$, if all SNRs were either emitting at the low or high end of observed luminosities for core-collapse SNRs.
The hard band emission is determined from observations of X-rays in the soft band scaled using a thermal bremsstrahlung spectrum with the temperature in Section~\ref{sec:spectral_properties}.
The green back-hatched (`\textbackslash') band shows the contribution from XRBs based on observations of the Galactic Centre and scaled by total stellar mass.
The red dashed line shows the hard component of unresolved emission from the old stellar component (mostly cataclysmic variables and active binaries) in the sphere of influence, as estimated in Ge~et.~al.~(2015).
The black point shows the X-ray luminosity observed for Sgr A*.}
\label{fig:luminosity_hard}
\end{figure}

These results from nearby galaxies suggest that, in the soft band, young SNRs are typically seen at $\lesssim 10^{37}~\mathrm{erg\,s}^{-1}$.
We therefore take this value as a conservative upper limit for our SNR luminosities.
A lower limit is more difficult to define, in part because there is no definitive boundary between the `adiabatic' and `radiative' stages, and similarly no break in the X-ray luminosities at any such point.
SNRs below $10^{35}~\mathrm{erg\,s}^{-1}$ in the soft band (close to the lower threshold of the ChASeM33 survey) are found to be middle-aged and of a well-evolved size.
Therefore, we take this luminosity as a lower limit for SNRs in this band.\footnote{Our upper limit is the more important prediction, as we are interested in high end of contaminating SNR luminosities in X-ray searches for quiescent SMBHs.}
These upper and lower limits define the hatched regions of SNR X-ray emission in Figs.~\ref{fig:luminosity_both}~and~\ref{fig:luminosity_hard} for the total SNR emission in the SOI.
This covers the extreme estimates of $L_\mathrm{X}$ if all of the $N_\mathrm{SNR}$ remnants were emitting at the very low or high ends of the expected luminosity from young core-collapse SNRs.

We note that our upper limit is conservative for at least two reasons. The first is that the ambient densities seen in SOI regions are larger than those seen in the typical ISM hosting the SNRs in these surveys. With a higher ambient density, the luminosity of the SNR is also expected to be higher.
The second is that, although the luminosities seen in the ChASeM33 survey (and as seen in the LMC and SMC) do not exceed $\sim 10^{37}~\mathrm{erg\,s}^{-1}$, luminosities at least an order of magnitude higher have been seen for very young SNRs in the Milky Way \citep[as noted in the aforementioned compilation of][]{Dwarkadas12}.

In our Galactic Centre, Sgr A East, has a present-day luminosity of $L_\mathrm{X,\,2-10\,keV} \approx 10^{35}~\mathrm{erg\,s}^{-1}$ \citep{Maeda02}.
This is the only well-known SNR near an SMBH, and it appears to be well into its adiabatic lifetime.
The luminosity for Sgr A East reassuringly lies between our upper and lower limits for the Milky Way value.

As we aim to compare with XRBs observed in the hard band, we convert between the soft ($0.35$--$2~\mathrm{keV}$) and hard ($2$--$8~\mathrm{keV}$) bands using a thermal bremsstrahlung spectrum with the temperature we determined in Section~\ref{sec:spectral_properties} (given in Fig.~\ref{fig:temperatures}).
We integrate the bremsstrahlung emissivity over these frequency bands ($\nu_\mathrm{\min}$ to $\nu_\mathrm{max}$) for a given temperature $\langle T' \rangle$, taking the free--free Gaunt factor to be approximately constant over these bands.
For a luminosity in the soft band, the corresponding luminosity in the hard band at $\langle T' \rangle$ is then given by the ratio
\begin{equation}
\frac{L_\mathrm{X, hard}}{L_\mathrm{X, soft}} = \frac{\exp{\left( \frac{- h \nu_\mathrm{min, hard}}{k_\mathrm{B} \langle T' \rangle} \right)} - \exp{\left( \frac{- h \nu_\mathrm{max, hard}}{k_\mathrm{B} \langle T' \rangle} \right) } }{\exp{\left( \frac{- h \nu_\mathrm{min, soft}}{k_\mathrm{B} \langle T' \rangle} \right)} - \exp{\left( \frac{- h \nu_\mathrm{max, soft}}{k_\mathrm{B} \langle T' \rangle} \right) } } .
\end{equation}

In Fig.~\ref{fig:luminosity_both}, we compare the SNR luminosity in the hard and soft bands by scaling the luminosity from a single SNR by $N_\mathrm{SNR}$, using the above conversion from soft to hard band luminosities.
This comparison makes it clear that around lower mass SMBHs, where the SNRs tend to evolve through to the radiative stage (and are therefore, on average, cooler), the expected emission favours the soft band.
On the other hand, SNRs around more massive SMBHs tend to be younger and hotter on average, and therefore the emission is stronger in the hard band.
Therefore, it is clear that SNRs may influence either soft or hard bands in SMBH searches, depending on the SMBH mass.

The sources of X-ray luminosity in the very dense and complex environments of galactic nuclei can be difficult to untangle.
As summarized in \cite{Ponti15}, the hard X-ray emission towards the Galactic Centre is substantially influenced by point sources \citep{Muno05}, and much of the hot thermal bremsstrahlung ($\sim 7.5~\mathrm{keV}$) emission seen in the region has been attributed to, at least at a $\sim 100~\mathrm{pc}$ scale, the integrated luminosities of unresolved sources \citep{Heard13a}.
The light from bright XRBs may additionally be scattered by the neighbouring ISM and molecular clouds, also at the scale of $\sim 100~\mathrm{pc}$ \citep{Sunyaev93, Molaro14}.

A bipolar outflow has also been observed about $14 \sim 20~\mathrm{pc}$ to either side of Sgr A*, with $L_\mathrm{X} \approx 10^{34}~\mathrm{erg\,s}^{-1}$ \citep{Morris03, Morris04, Markoff10, Heard13b, Ponti15}.
This has been attributed to either shock-heated winds from massive stars or tidal disruption events \citep{Heard13b}, or as another possible SNR due to the recently revealed presence of a shock at the lobe boundaries \citep[with a possible $2 \sim 4~\mathrm{keV}$ component;][]{Ponti15}.

Here, we restrict our comparison to two other possible X-ray sources that are of interest as contaminants in X-ray searches for quiescent SMBHs: resolved XRBs and unresolved emission from the old stellar component of the nucleus.
\cite{Muno05} have reported the detection of four XRBs within only $1~\mathrm{pc}$ of the Galactic Centre.
To characterize these as XRBs, the selection of sources was restricted to those with large outbursts to distinguish them from other, consistently bright point sources.
These sources had peak emissions between $L_\mathrm{X,\,2-8\,keV} \approx 10^{33}$ and $10^{35}~\mathrm{erg\,s}^{-1}$, which is in fact a peculiar range that is between typical values of quiescent and outbursting XRBs; this also makes it unclear whether these are high- or low-mass XRBs \citep{Campana98, Muno05}.

Fig.~\ref{fig:luminosity_hard} shows an estimate of the combined emission from known point-source, active XRBs at a given time, based on these observations.
We use the range of peak luminosities of the four active XRBs seen in the inner $\sim 1~\mathrm{pc}$ of the Galactic Centre ($10^{33} \sim 10^{35}~\mathrm{erg\,s}^{-1}$), which is represented as a green back-hatched (`\textbackslash') band.
As these four XRBs did in fact vary in luminosity over the observed \textit{Chandra} period, taking their peak luminosities for the band on Fig.~\ref{fig:luminosity_hard} will therefore likely be a conservatively high estimate of the total luminosity.
This estimate also implicitly incorporates the remainder of the XRB population as being in quiescence and below the detection threshold at a given time.
The X-ray emission for other nuclei is calculated by taking the same ratio of confirmed XRBs to total stellar mass (which scales linearly with $\Mbh$) as that observed in the Galactic Centre.

Unresolved X-ray emission also originates from the old stellar population in the region, and is contributed to mainly by cataclysmic variables (CVs) and active binaries (ABs).
The associated luminosity has been found to roughly scale with the stellar mass in observations of the Local Group \citep{Revnivtsev06, Revnivtsev09, Ge15}.
We take the hard-band relation $L_\mathrm{X,\,2-8\,keV} \approx 10^{27}~\mathrm{erg\,s}^{-1}~M_*/\Msun$ given in \cite{Ge15}, using the scaling $M_* \approx 2 \Mbh$ in the SOI (Section \ref{sec:autarkic_model}).
This estimate is given in Fig.~\ref{fig:luminosity_hard} as the red dashed line.

It is clear from Fig.~\ref{fig:luminosity_hard} that if nuclei scale similarly with the Galactic Centre, then it is possible for SNRs to compete with the X-ray emission from XRB point sources as well as the unresolved X-ray emission in the hard band.
Furthermore, the emission from all of these source types is more luminous than the current X-ray luminosity of Sgr A* itself (a few $10^{33}~\mathrm{erg\,s}^{-1}$).
Therefore, for other SMBHs of similar Eddington ratios and $L_\mathrm{X}/L_\mathrm{Edd}$ as Sgr A*, the emission from the central engine can be overwhelmed by contamination from both XRBs and young SNRs.

\subsection{Detectability} \label{sec:detectability}
\begin{figure*}
\captionsetup[subfigure]{labelformat=empty}
\centering
\subfloat[]{\includegraphics[width=\columnwidth]{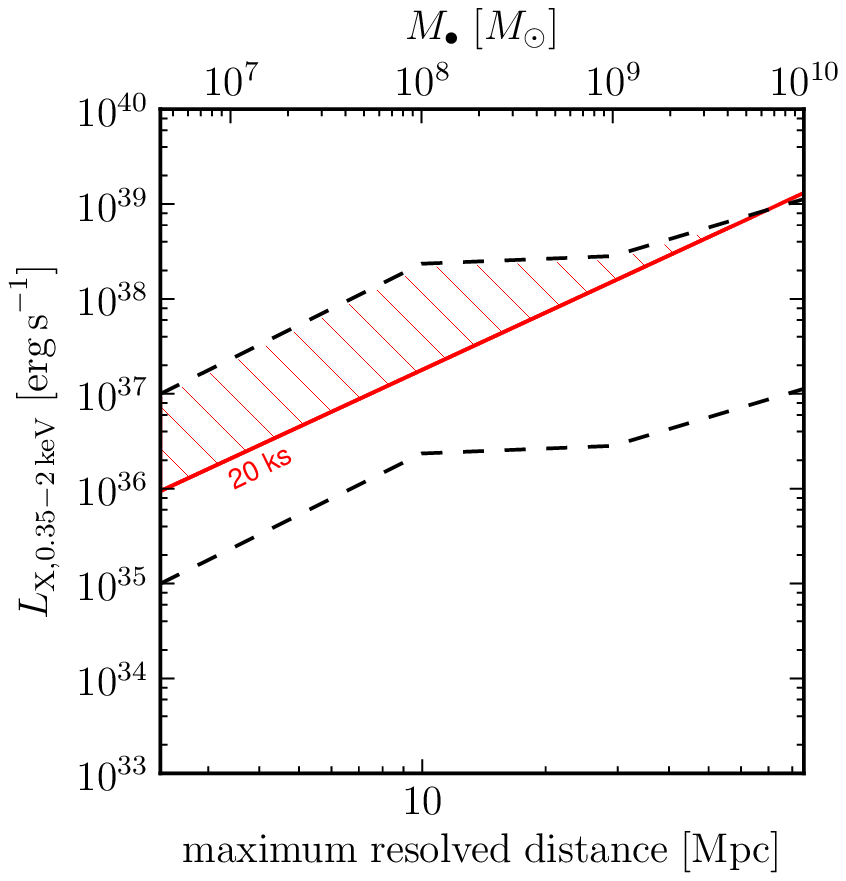}}
\subfloat[]{\includegraphics[width=\columnwidth]{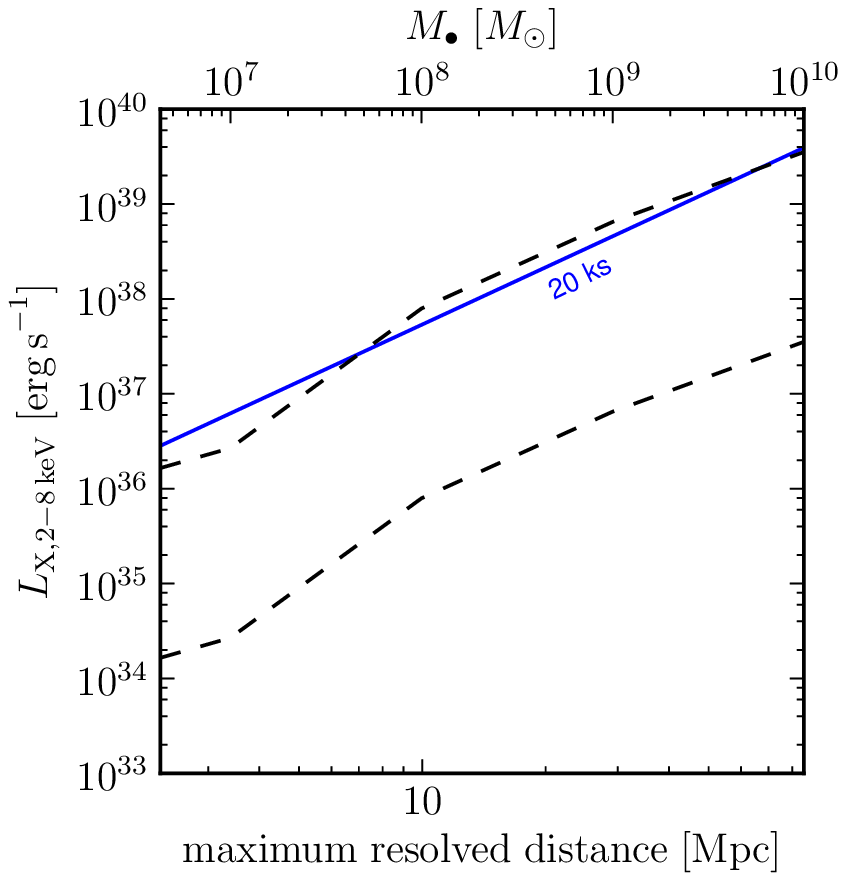}}
\caption{Observable limits of the combined X-ray luminosities from young SNRs as a function of SMBH mass (upper axes) or maximum resolved distance (lower axes).
The maximum resolved distance is that at which the SOI of the SMBH is within the core of the PSF of \textit{Chandra}.
The left-hand panel shows luminosities in the soft X-ray ($0.5$--$2~\mathrm{keV}$), while the right-hand panel shows hard luminosities ($2$--$8~\mathrm{keV}$).
SNR luminosity limits are the dashed black lines (shown also in Fig.~\ref{fig:luminosity_hard}).
Flux-limited detection thresholds from \textit{Chandra} exposure times of $20~\mathrm{ks}$ for galaxies at the distance limit are given as solid lines.
The shaded region in the left-hand panel highlights the observable range of luminosities in the soft band.}
\label{fig:luminosity_obs}
\end{figure*}

We now consider whether these predictions can be observed, exploiting the high spatial resolution ($0.49~\mathrm{arcsec}$) of the \textit{Chandra} satellite.
Fig.~\ref{fig:luminosity_obs} shows the expected soft ($0.5$--$2~\mathrm{keV}$) and hard ($2$--$8~\mathrm{keV}$) X-ray luminosity as a function of the black hole mass.
The upper and lower limits of the SNR luminosities in Fig.~\ref{fig:luminosity_obs} are found in Section~\ref{sec:x-ray_luminosity} and are the same as those plotted on Fig.~\ref{fig:luminosity_both}.
The solid lines of Fig.~\ref{fig:luminosity_obs} show the range of luminosities that can be detected by {\it Chandra}-ACIS-S with 20 ks exposures (based on flux limits of $3 \times 10^{-15}$\,erg\,cm$^{-2}$\,s$^{-1}$ and $9 \times 10^{-15}$\,erg\,cm$^{-2}$\,s$^{-1}$ for a $10~\mathrm{ks}$ exposure).
The lower horizontal axes are given in terms of the maximum resolvable distance of the SOI.
This is the distance at which the (diameter of the) SOI of the SMBH is just within the core of the point-spread function of \textit{Chandra}.
For example, the SOI of a $10^8~\Msun$ SMBH is resolvable at any distance below $\sim 10~\mathrm{Mpc}$; and, at $10~\mathrm{Mpc}$, the lower limit of detectable soft X-ray emission is given by the red line.
At any distance less than $10~\mathrm{Mpc}$, this detection threshold drops and so fainter emission from SNRs would be detectable.

From Fig.~\ref{fig:luminosity_obs}, it is evident that, even at the maximum resolved distance, the upper limit of SNR emission is detectable as it is above the $20~\mathrm{ks}$ threshold in the soft band through most of the $\Mbh$ range (although the low end of possible X-ray luminosities is possibly not within the detection threshold of \textit{Chandra}).
In the hard band, the \textit{Chandra} $20~\mathrm{ks}$ threshold follows very closely the high limit of predicted SNR emission.
This suggests that, for galaxies at the maximum resolved distance, the emission would not be detectable in the hard band.
Again, however, if a galaxy is closer than this maximum distance, the threshold marked by the blue solid line will drop; therefore, hard X-ray luminosities are still potentially detectable for more nearby galaxies.

\begin{figure}
\centering
\includegraphics[width=\columnwidth]{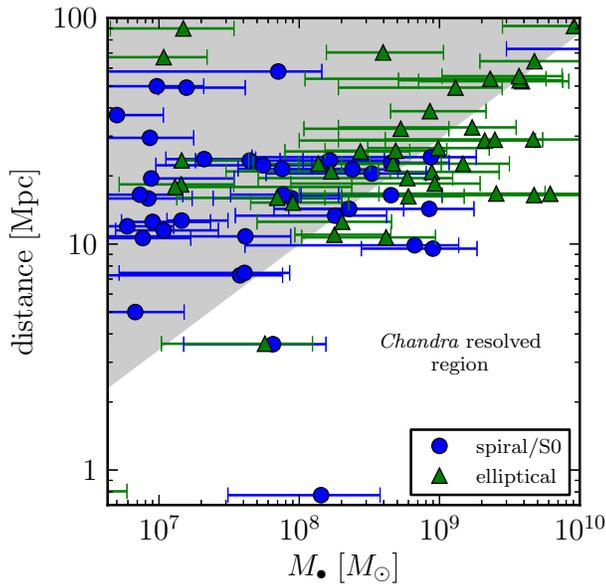}
\caption{Distances and masses (with errors) of known SMBHs over the same mass range considered in this paper.
Data were taken from tables in the review of Kormendy \& Ho (2013). 
Masses were estimated primarily from stellar dynamics, with, in some cases, measurements of gas motions near the SMBH.
Blue circles represent spiral or lenticular (S0 Hubble type) galaxies, while green triangles represent elliptical galaxies.
The grey region shows the combinations of $\Mbh$ and distance for which the angular size of the SOI is within the core of the \textit{Chandra} point-spread function and the SOI is unresolved.
The white region therefore shows the conditions for which the SOI is resolved.}
\label{fig:known_SMBHs}
\end{figure}

To compare this with the number of SMBHs known at these distances, in Fig.~\ref{fig:known_SMBHs} we plot the distances and masses of well-established SMBHs constructed from Tables 2 and 3 in the review of \cite{Kormendy13} and the references therein.
We show in the grey region the cases where the SOI of the SMBH is not resolved by \textit{Chandra}.
As for the upper axes in Fig.~\ref{fig:luminosity_obs}, the limit between the grey and white area is determined by the distance at which the SOI of the SMBH equals in size the core of the point-spread function of \textit{Chandra}.

Although the majority of known SMBH SOIs lie in this unresolvable region, a large fraction ($\sim 30$) of the candidates stand out and may be targets to compare with our predictions.
Many potential candidates within the axis limits of Fig.~\ref{fig:known_SMBHs} are members of the Virgo and Fornax clusters \citep{Jordan07, Ferrarese12}. 
Most of the resolvable SOIs belong to SMBHs with masses $10^8 \sim 10^9~\Msun$, many of which lie well within the maximum resolved distance.

We therefore conclude that our predictions and thus the ansatz of self-regulation and self-similarity for quiescent galactic centres may be testable, currently, for a reasonable population of galaxies.\footnote{One additional hindrance to observing nuclear sources is the inclination of late-type galaxies to our line of sight. The nuclei of edge-on galaxies are potentially more contaminated by unresolved X-ray point sources and hot, X-ray emitting gas.}
An obvious next step would be to perform a systematic search of \textit{Chandra} archives for specific examples, but this is beyond the scope of the current work.

\section{The sphere of influence SFR} \label{sec:SFR}
\begin{figure}
\centering
\includegraphics[width=1.08\columnwidth]{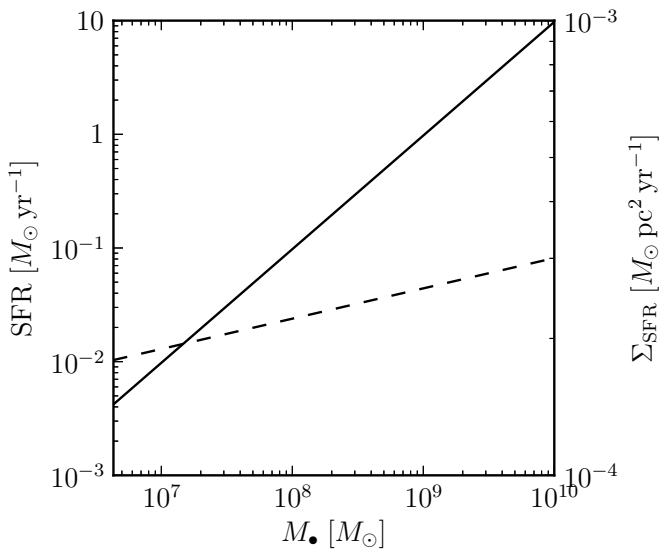}
\caption{Total (solid line; left-hand axis) and surface density (dashed line; right-hand axis) SFRs within the SMBH sphere of influence as a function of SMBH mass.}
\label{fig:SFR}
\end{figure}
Since massive stars trace the star formation history of a region, our previous results also allow us to estimate the SFR in the SOI as a function of the black hole mass.
In steady state, the supernova rate, $\mathcal{R}_\mathrm{SN}$, is equal to the rate of formation of new stars:
\begin{align}\label{eq:supernova_rate_SFR}
\mathcal{R}_\mathrm{SN} = \frac{N_\mathrm{cc} }{\langle t_*(M>8) \rangle} &= \mathrm{SFR} \, \frac{\int_{8 \Msun}^{50~\Msun} {\varphi(M)} \, \ud M }{\int_{0.1~\Msun}^{100~\Msun} {M \varphi(M)} \, \ud M } \nonumber \\ 
                       &= 1.2 \times 10^{-2}\,\mathrm{yr}^{-1} \left( \frac{\mathrm{SFR}}{ \Msun\,\mathrm{yr}^{-1}} \right),  
\end{align}
where `SFR' is the total SFR spread over our fiducial IMF (equation \ref{eq:imf}), and
\begin{equation}
\frac{N_\mathrm{cc}}{\langle t_*(M>8) \rangle} \approx 5 \times 10^{-5} \left( \frac{\Mbh}{4.3 \times 10^6~\Msun }\right)~\mathrm{yr}^{-1},
\end{equation}
combining equations~(\ref{eq:t_star}) and (\ref{eq:number_cc}).
This allows us to write the SFR as a function of the black hole mass:
\begin{equation} \label{eq:SFR}
\mathrm{SFR} \approx 4 \times 10^{-3} \left( \frac{\Mbh}{4.3 \times 10^6~\Msun} \right)~\mathrm{M}_{\odot}\,\mathrm{yr}^{-1} .
\end{equation}

The total SFR as a function of $\Mbh$ is shown as the solid line (left-hand axis) in Fig.~\ref{fig:SFR}.
This corresponds to an SFR density averaged over the whole SOI, $\Sigma_{\rm SFR}$, 
that stays approximately constant around $10^{-4}~\Msun\,\mathrm{yr}^{-1}\,\mathrm{pc}^{-2}$ in the whole 
mass range of interest (dashed line, right-hand axis of Fig.~\ref{fig:SFR}).
Making the IMF more top-heavy ($\alpha = 0.45$) does not change the multiplicative factor in the right-hand side of equation~(\ref{eq:supernova_rate_SFR}) within the given precision ($1.2 \times 10^{-2}$), and so the predicted SFR is unaffected.

Few observations of stellar populations at the scale of quiescent SOI are available for comparison with these predictions, not only due to the challenge of resolving parsec-scale properties but also due to obscuration of the nuclear star clusters.
We gather some observations below.

For the SOI of the best-studied SMBH, Sgr A*, equation~(\ref{eq:SFR}) predicts an SFR of $4 \times 10^{-3}~\mathrm{M}_{\odot}\,\mathrm{yr}^{-1}$, equivalent to an SFR per unit area of $\Sigma_\mathrm{SFR} = \mathrm{SFR} / \left(\uppi R_\mathrm{SOI}^2 \right) \approx 2 \times 10^{-4}~\Msun\,\mathrm{yr}^{-1}\,\mathrm{pc}^{-2}$.
This is in agreement with the observationally inferred SFR of $7 \times 10^{-4}~\Msun\,\mathrm{yr}^{-1}\,\mathrm{pc}^{-2}$ in the innermost $1.2~\mathrm{pc}$ (which we expect to have a higher SFR density than the outer sphere of influence), in the last $\sim 10^7~\mathrm{yr}$ \citep{Pfuhl11}.

A sharp increase in SFR is seen with decreasing distance from the centre of many nearby spiral galaxies \citep{Schruba11}, even if the SOI of the putative SMBH is not resolved.
Extrapolating the values of SFR towards the centre of the Milky Way and NGC 6946, shown in \cite{Kennicutt12}, gives values in line with those predicted here.
In small-bulged, late-type spirals, \cite{Walcher06} can directly resolve the nuclear star clusters within the SOI, because they are not obscured by the presence of a massive bulge.
Extending the results to lower mass, the values we find are reasonable for these galaxies.
We compare with \cite{Walcher06} by using their SFR calculated over the most recent $10^8~\mathrm{yr}$.
This gives $\Sigma_\mathrm{SFR} \approx 6 \times 10^{-5}~\Msun\,\mathrm{yr}^{-1}\,\mathrm{pc}^{-2}$, for a mean $\Mbh \approx 10^{5}~\Msun$, where we use the $\Mbh$ estimates in \cite{Neumayer12}, and we take a radius equal to the mean effective radius of $3.5~\mathrm{pc}$ from \cite{Boker04}.

Despite the limited observations of SNRs in the Galactic Centre available to anchor our results, these comparisons give support to our autarkic scenario for quiescent galactic nuclei.
Ideally, a combination of X-ray observations of SNRs and nuclear SFRs would be needed to refine these predictions.

\section{Discussion and Conclusions} \label{sec:conclusion}
Quiescent galactic nuclei such as that of the Milky Way are frequently seen to harbour massive stars.
We have demonstrated, elaborating on work presented in Paper I, that their presence can be exploited to gain insights into these common, but not well understood, environments.

Our model for SNR evolution, developed in Paper I, can be applied to a diverse range of descriptions of the regions near SMBHs.
However, in particular, we have chosen an `autarkic' framework that is consistent with observations of the Galactic Centre.
This describes quiescent SMBH environments that are self-regulating and uninfluenced by any inflows of material beyond the sphere of influence.
The gas in the accretion flow is supplied by stellar winds, which also provides part of the material from which new stars are formed.
In this model, we take the total rate of star formation, and therefore number of massive stars in steady state, to scale with the total stellar mass in the sphere of influence.
As a consequence, the accretion rate is the same fraction of Eddington ($\dot{M}/\dot{M}_\mathrm{Edd} \approx 10^{-5}$) as that of Sgr A* in the Galactic Centre.

For SNe exploding in such environments, our dynamic modelling predicts the `adiabatic' lifetimes and therefore total number of core-collapse SNe seen at one time.
We find $1 \sim 10^2$ SNRs in the sphere of influence of SMBHs over the mass range $10^{6}$ to $10^{10}~\Msun$.
As the SMBH mass increases, the reduced lifetime of SNRs competes with the increase in core-collapse progenitors in the region, resulting in a sublinear increase of the observed number of SNRs.

In galactic nuclei beyond the Local Group, for which the resolution of individual SNRs may be more challenging, the presence of hot SNRs can affect the total X-ray emission from the sphere of influence.
Therefore, we use the total number of SNRs to estimate the total X-ray emission expected from these regions.
One caveat, noted also in Paper I, is that the Kompaneets approximation may overestimate the `adiabatic' lifetimes of SNRs.
Correcting for this may reduce the X-ray luminosities predicted here, in particular for low-mass SMBHs (which have, on average, longer lived SNRs).

We find that, for nuclei with properties like that of our Galactic Centre, core-collapse SNRs can compete with the emission from X-ray binaries as well as unresolved sources, and can potentially outshine the emission from the central engine itself.
This is indeed what is observed for the known X-ray sources in the Galactic Centre, where different X-ray sources be more easily distinguished; the X-ray luminosity within the Sgr A East shell \citep[$\sim 10^{35}~\mathrm{erg\,s}^{-1}$;][]{Maeda02} is higher than that observed from Sgr A* \citep[a few $10^{33}~\mathrm{erg\,s}^{-1}$;][]{Baganoff03, Wang13}.
We predict that this emission could be detectable, particularly in the soft band, out to the maximum distance at which the SMBH sphere of influence is resolved by \textit{Chandra}.
Though beyond the scope of the current work, a natural follow-up would be to examine \textit{Chandra} data for specific galaxies.

Knowledge of SNR lifetimes can be used to estimate the SFR and core-collapse progenitor numbers in these environments.
For a Milky Way-type galaxy, our estimated SFR density of a few $10^{-4}\,\Msun\,\mathrm{yr}^{-1}\,\mathrm{pc}^{-2}$ is in good agreement with other approaches \citep{Pfuhl11}.
This corresponds to a core-collapse progenitor population of $N_\mathrm{cc} = 500 \sim 1000$ within the sphere of influence.
The SFR obtained for other galactic nuclei shows concordance with available data on SFR in nuclear clusters.
\cite{Kennicutt12} show the SFR as a function of radius for the specific examples of the Milky Way and NGC 6946, which, at small radii, tend towards the values we predict.
Ideally, a combination of X-ray measurements and estimates of SFRs can then be used to compare with our predictions and understand at what extent quiescent nuclei are autarkic.

The gas models in this paper were refined by adding molecular tori around the scale of the black hole sphere of influence.
Our results with the molecular tori show that they do not generally have a substantial effect on the lifetimes of SNRs or on the overall morphology of the SNR.
The minimal impact of the torus on SNR expansion in the Galactic Centre suggests that large-scale asymmetries, such as those seen in Sgr A East or the $20~\mathrm{pc}$ lobes, are not due to dynamical confinement from the molecular torus alone.

One potential improvement to our current approach relates to the X-ray luminosity predictions, which rely on data and are not calculated \textit{ab initio}.
Although the use of observations allowed us to make qualitative inferences on the importance of SNRs, more detailed predictions 
would require a more accurate modelling of the radiative processes within the SNR.
This prevents us from quantifying more conclusively the contribution to X-ray of SNRs (Fig.~\ref{fig:luminosity_obs}), or to draw conclusions on the feasibility to use X-ray observations to constrain the gaseous ambient medium gradient.
This possibility of using SNRs to constrain the gaseous environment can be seen, for example, in the differences of a factor of a few in the predicted $N_\mathrm{SNR}$ around a $10^9~\Msun$ SMBH in Fig.~\ref{fig:lifetimes_and_numbers}.
Constraining the inner gaseous medium bears the exciting promise of pinning down the physics that describes radiatively inefficient accretion flows.

In this work, we have not taken into account additional processes over short time-scales that may affect the accretion rate and luminosity of the SMBH.
Temporary increases in accretion rates can occur from the input of small amounts of stellar or gaseous mass from outside the sphere of influence.
For example, stellar tidal disruption events like that of \textit{Swift}~J1644+57 produce SMBH flaring observed in radio through to X-rays \citep{Komossa02, Burrows11}.
An inwards deposition of molecular gas, tidally disrupted into a disc, might also be responsible for the formation of some of the young stars around SMBHs \citep{Levin03, Paumard06}.
We also emphasize that, although SNRs tend to sweep gas out of their environment, explosions near the SMBH may also, at a smaller scale, deposit some material near the SMBH, enhancing accretion and causing temporary flaring.

With respect to outflows, a simple estimate of their importance can be made by comparing the gravitational binding energy of the SMBH-gas systems with that of the total supernova energy.
Doing this calculation, we find that the supernova energy is larger than the binding energy at low $\Mbh$, and that the gravitational binding energy of the gas grows faster than the supernova energy input, but the two values only become comparable by large SMBH masses ($\sim 10^{9}~\Msun$).
If the supernova energy is efficiently deposited into kinetics of the nuclear gas, then outflows due to SNe in quiescent nuclei may be important around lower mass SMBHs.
Large-scale expulsion of gas may temporarily reduce the rate of accretion from stellar winds; however, as has been shown for lower mass SMBHs like Sgr A* (Section~\ref{sec:total_numbers}, the gas refilling time-scale is shorter than the supernova rate.
Such core-collapse-induced outflows from low-mass SMBHs are a potentially interesting topic of future work.

Finally, our model is anchored to Galactic Centre observations an influenced by uncertainties on the number of current SNRs in the adiabatic phase.
These uncertainties certainly propagate through our predictions.
With the constant monitoring of the Galactic Centre, as seen in the recent results of \cite{Ponti15}, this model will no doubt become better informed.
 
We have shown that this autarkic model applied to young stars, gas, and the SNRs amongst them, is a promising framework to understand how these nuclei function and evolve.
The predictions of our model can be tested and refined by X-ray observations and SFR estimates of quiescent galactic nuclei.
 
\section*{Acknowledgements}
This work was supported by the Netherlands Research Council (NWO grant numbers 612.071.305 [LGM] and 639.073.803 [VICI]) and by the Netherlands Research School for Astronomy (NOVA).
The authors thank the anonymous referee for very helpful comments on the manuscript.
We also gratefully acknowledge Tsvi Piran for feedback that shaped an early version of this work, and we wish to thank Elena Gallo and Gabriele Ponti for helpful discussions.

\bibliographystyle{bib/mn2e}
\footnotesize{

}

\label{lastpage}

\end{document}